\documentclass[conference]{IEEEtran}
\usepackage{amsmath,amssymb,epsfig}
\usepackage{graphicx}
\usepackage{flushend}
\usepackage[utf8]{inputenc} 
\usepackage[T1]{fontenc}    
\PassOptionsToPackage{hyphens}{url}\usepackage{hyperref}
\usepackage{url}            
\usepackage{booktabs}       
\usepackage{nicefrac}       
\usepackage{microtype}      
\usepackage{xspace}
\usepackage{tikz}

\newcommand{\figframe}[1]{\fbox{#1}}

\newcommand{\etal}[0]{\textit{et al.}\xspace}
\newcommand{\tseqpre}{\textsc{EQPred}\xspace}
\newcommand*\circled[1]{
  \tikz[baseline=(char.base)]{\node[shape=circle,fill,inner sep=0,minimum size=1.2em] (char) {
              \textcolor{white}{#1}};}}

\newcounter{modelcounter}
\newcommand{\modelnum}[1]{\stepcounter{modelcounter}\circled{\themodelcounter} {#1}}

\begin{document}

\title{Spatiotemporal Pattern Mining for Nowcasting Extreme Earthquakes in Southern California}

\author{
  \IEEEauthorblockN{Bo Feng}
  \IEEEauthorblockA{Intelligent Systems Engineering\\
    Indiana University Bloomington\\
    Bloomington, Indiana, USA\\
    Email: fengbo@iu.edu}
  \and
  \IEEEauthorblockN{Geoffrey C. Fox}
  \IEEEauthorblockA{Intelligent Systems Engineering\\
    Indiana University Bloomington\\
    Bloomington, Indiana, USA\\
    Email: gcf@indiana.edu}
}

\maketitle

\pagenumbering{arabic}
\pagestyle{plain}

\begin{abstract}
    Geoscience and seismology have utilized the most advanced technologies and equipment to monitor seismic events globally from the past few decades. With the enormous amount of data, modern GPU-powered deep learning presents a promising approach to analyze data and discover patterns. In recent years, there are plenty of successful deep learning models for picking seismic waves. However, forecasting extreme earthquakes, which can cause disasters, is still an underdeveloped topic in history.  Relevant research in spatiotemporal dynamics mining and forecasting has revealed some successful predictions, a crucial topic in many scientific research fields. Most studies of them have many successful applications of using deep neural networks. In Geology and Earth science studies, earthquake prediction is one of the world's most challenging problems, about which cutting-edge deep learning technologies may help discover some valuable patterns. In this project, we propose a deep learning modeling approach, namely \tseqpre, to mine spatiotemporal patterns from data to nowcast extreme earthquakes by discovering visual dynamics in regional coarse-grained spatial grids over time. In this modeling approach, we use synthetic deep learning neural networks with domain knowledge in geoscience and seismology to exploit earthquake patterns for prediction using convolutional long short-term memory neural networks. Our experiments show a strong correlation between location prediction and magnitude prediction for earthquakes in Southern California. Ablation studies and visualization validate the effectiveness of the proposed modeling method.
\end{abstract}

\begin{IEEEkeywords}
    Spatiotemporal, Convolution, Recurrent Neural Network, LSTM, Temporal Convolution, Nowcasting
\end{IEEEkeywords}

\IEEEpeerreviewmaketitle

\section{Introduction}
\label{sec:intro}

Spatial and temporal attributes have played an essential role in addressing scientific issues mathematically and statistically with large volumes of data in real problems. A worldwide team of scientists studied the published datasets from The WorldPop project (www.worldpop.org) for discovering the spatiotemporal pattern of population in China from 1990 to 2010~\cite{gaughan_spatiotemporal_2016}. For modern Geoscience, spatiotemporal modeling has been studied for a long time. In this book~\cite{christakos_modern_2000}, authors summarized some initial efforts by utilizing spatiotemporal features for scientific interpretation and prediction.

Tradition machine learning algorithms like the support vector machine (SVM) and decision trees perform well on small datasets. Optimization methods such as stochastic gradient descent (SGD) enable the deep learning algorithms can be trained in small batches for extensive data without sacrificing model performance.
Over the past few decades, large volumes of data have been collected by the seismological community. This drives high demand for seismology data processing and analysis, providing opportunities to predict future dynamics from history. Spatiotemporal forecasting is an important topic in many scientific research fields, in which there are a plethora of successful applications. Recent studies using deep neural networks have shown various successful applications, including car traffic forecasting~\cite{li_diffusion_2017}, ride-hailing forecasting~\cite{zhu_deep_2017}, rain/weather forecasting~\cite{wang_deep_2020}, etc.

\begin{figure}
    \begin{minipage}[b]{.49\linewidth}
        \centering
        \centerline{\includegraphics[width=3.9cm, height=2.6cm]{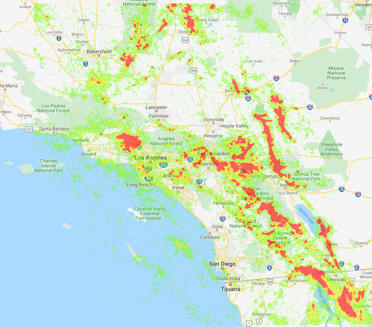}}
        \centerline{(a)}\medskip
        \centerline{\figframe{\includegraphics[width=3.9cm, height=2.6cm]{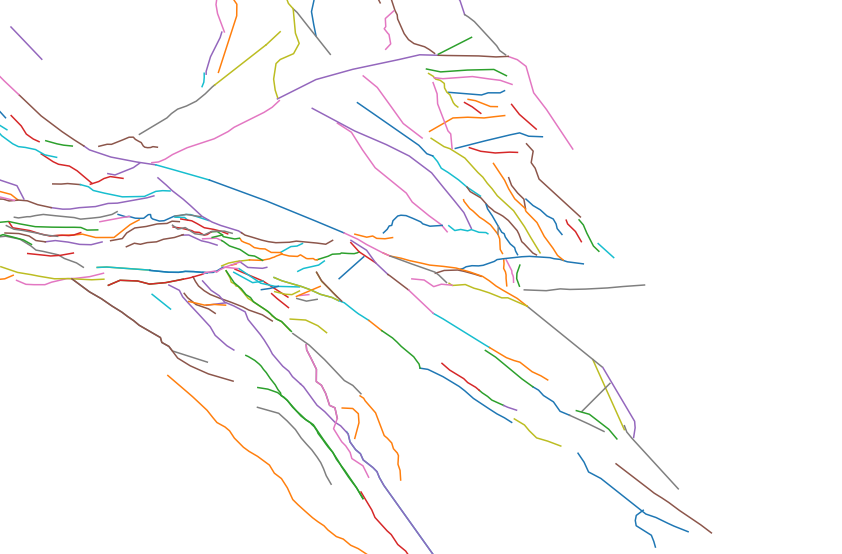}}}
        \centerline{(c)}\medskip
    \end{minipage}
    \hfill
    \begin{minipage}[b]{0.49\linewidth}
        \centering
        \centerline{\includegraphics[width=3.9cm, height=2.6cm]{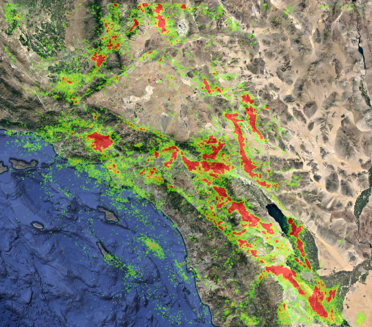}}
        \centerline{(b)}\medskip
        \centerline{\includegraphics[width=3.9cm, height=2.6cm]{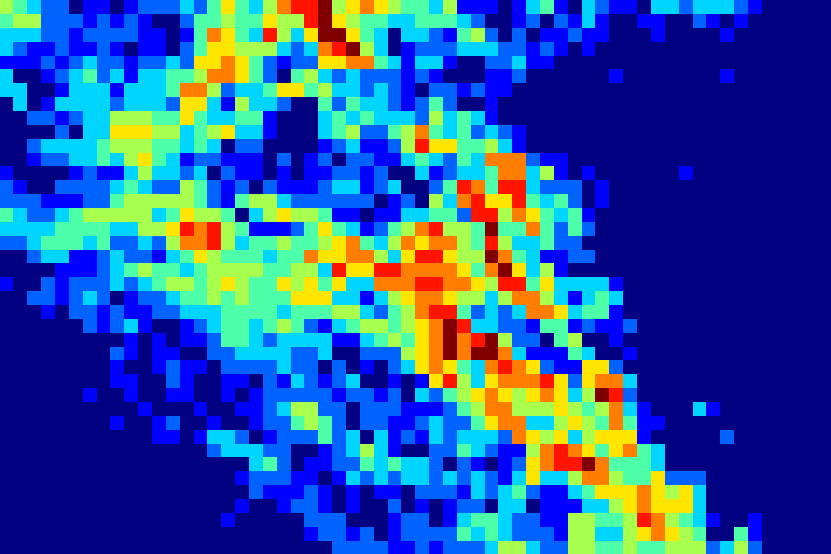}}
        \centerline{(d)}\medskip
    \end{minipage}
    \caption{Dataset overview of earthquakes in Southern California.  (a) Earthquake events mapped on Maps. (b) Earthquake events mapped on satellite images. (c) Earthquake fault lines in the same region. (d) A heatmap view of events.}
    \label{fig:dataset}
\end{figure}

Recent studies in deep neural networks have many successful applications of using deep learning for spatiotemporal forecasting. The goal of spatiotemporal forecasting is to predict what and when the next event will happen. This is a task that includes two orthogonal sub-tasks: forecasting its spatial dependencies and temporal dependencies. However, this is a nontrivial task due to the high dimension features of time series sequences and building models that can work well for some specific problems can also be very vague.

Earthquakes are caused by the sudden release of energy from seismic waves~\cite{bakun_estimating_1997,mousavi_earthquake_2020}. However, this involves the movement of ground plates via a stochastic process, which makes Earthquake forecasting is a worldwide challenging problem. Scientists around the world have built an enormous number of detectors for picking up earthquake signals. It is a general belief that earthquakes are predictable under some assumption that quakes are formed underneath the Earth are accumulated stresses in a gradual process over a long time. In this case, it would be possible to predict earthquake shocks for future activities of quakes by learning patterns from historical seismic events.

Conventionally, earthquakes are located through a process of detecting signals, picking up arrival time, and estimating epicenters of events using a velocity model. Efforts have been made to filter P-waves and S-waves from the original waveform signals of earthquakes and seismic noise~\cite{mousavi_earthquake_2020}. In this project, our goal is to utilize the preprocessed seismic signals forming epicenters (location labels) to forecast the probabilities of the subsequent earthquakes in an area.

Earthquake forecasting consists of three major tasks in machine learning. The first task is to predict when the next seismic event will happen in a specific region. The second task is to predict whether or not the next seismic event will come. The third task is to predict the level of magnitude of the upcoming seismic events to predict major shocks.

Deep learning neural networks have presented a widely successful approach to capture spatial-temporal dependencies of problems to achieve accurate forecasting results. Convolutional neural networks have achieved convinced success in computer vision, image object recognition, etc~\cite{lecun_deep_2015}. Here we test the hypothesis that earthquake patterns can be perceived by learning historical seismic events. However, epicenters' prediction is learned from annotated seismograms. Due to the uncertainties of earthquakes, even the ground truth labels that are annotated by domain experts may be biased. Locations and magnitude of epicenters are maybe adjusted after the seismic event happened a long while.

In this project, we propose joint modeling of using a self supervised autoencoder (AE) and temporal convolutional neural networks (TCN)~\cite{oord_wavenet_2016} for earthquake prediction by modeling spatiotemporal dependencies in Southern California.
Additionally, \tseqpre comprehensively improves the autoencoder and TCN by incorporating skip connections and local temporal attention mechanisms.
Compared to conventional recurrent neural networks or a single model, our joint modeling presents some advantages in predicting major shocks in the area of study. In summary:
\begin{itemize}
  \item We study the earthquake dataset for Southern California and reconstruct the time series events into a sequence of 2D images.
  \item We model the spatiotemporal dependencies of earthquakes in Southern California with an improved autoencoder and TCN neural networks and show some preliminary but promising results for nowcasting events in contrast to 11 baseline models.
\end{itemize}

This paper is organized as following: Section~\ref{sec:rel} reviews related work in four aspects. In Section~\ref{sec:tseqpredictor}, we propose a spatiotemporal modeling approach, namely \tseqpre, and illustrate the how spatial and temporal dynamics are modeled theoretically. In Section~\ref{sec:exp}, we show the detailed implementation of \tseqpre, run analysis on the dataset, and evaluate the effectiveness. Finally, we conclude in Section~\ref{sec:con} with the discussion of the limitations and future research directions.

\section{Related Work}\label{sec:rel}

Four subjects are related to this project: 1) predicting epicenters only with advanced machine learning techniques; 2) spatiotemporal modeling in a broad range of applications; 3) advanced dynamic pattern mining and prediction in visual applications; 4) extreme event prediction in other areas of research.

\subsection{Convolutional Methods in Predicting Epicenters}
Estimating and predicting the epicenters of earthquakes has a long history. Scientists from Geophysics, Geology, and Seismology have developed various tools and analytical functions to predict epicenters from datasets. In 1997, Bakun and Wentworth suggested using Modified Mercalli intensity datasets for Southern California earthquakes to bound the epicenter regions and magnitudes~\cite{bakun_estimating_1997}. In 1998, Pulinets proposed predicting epicenters of strong earthquakes with the help of satellite-sounding systems. Scientists from Greece had illustrated a successful project which predicted the prominent aspects of earthquakes using seismic electric signals~\cite{pulinets_strong_1998}. Recently, Guangmeng \etal attempted to predict earthquakes with satellite cloud images and revealed some possibilities of predicting earthquakes using geophysics data~\cite{guangmeng_three_2013}. Zakaria \etal presented their work of predicting epicenters by monitoring precursors, such as crustal deformation anomalies and thermal anomalies, with remote sensing techniques~\cite{alizadeh_zakaria_possibility_2020}. These studies either used only too little data or too simple analytical models.

\subsection{Spatiotemporal Dynamics and Generative Models}
Most recently, it is a prevailing method to make predictions by modeling the spatiotemporal dynamics for domain science problems. This is because large volumes of data are increasingly collected in the vast majority of domains including, social science, epidemiology, transportation, and geoscience. Cui \etal proposed to use graph convolutional long short-term memory neural networks to predict traffic via capturing spatial dynamics from the car traffic patterns~\cite{cui_traffic_2019}. Li \etal utilized a seq2seq neural network architecture to capture spatial and temporal dependencies for traffic forecasting by incorporating a diffusion filter in convolutional recurrent layers~\cite{li_diffusion_2017}. FUNNEL was a project proposed by Matsubara~\cite{matsubara_funnel_2014}. It was designed to use an analytical model and a fitting algorithm for discovering spatial-temporal patterns of epidemiological data.

\subsection{Visual Pattern Prediction}
Lotter \etal presented a model to predict video frames with deep predictive coding networks~\cite{lotter_deep_2017}, which was based on the ConvLSTM2D network module with specific top-down states updating algorithm. \cite{finn_unsupervised_2016} is another example in predicting video frames. The authors of this work presented the effectiveness of modeling object motion via predicting future object pixels. For example, a ball moves and a block falls.
These models are successful for predicting contiguous and dense image frames, whereases the earthquake data are very sparse, the extreme shocks are very rare in terms of probability.

\subsection{Extreme rare event prediction}
Laptev \etal~\cite{laptev_time-series_nodate} proposed their modeling to predict rare trip demands for ride-hailing service. In that paper, they built an end-to-end architecture using joint modeling by combining LSTM autoencoder and LSTM predictor networks. They showed their forecasting capability on a Uber's public dataset.
Geng \etal~\cite{geng_spatiotemporal_2019} proposed another model to forecast the ride-hailing demand using graph-based recurrent neural networks, in which graphs are defined by road networks with Euclidean and non-Euclidean distances.
We compared this approach with our \tseqpre, a detailed discussion of which is in Section~\ref{sec:perf-analysis}.

\section[TSeqPredictor]{\tseqpre Modeling}
\label{sec:tseqpredictor}

Earthquakes come with an epicenter which is a point at the surface on Earth, and the mainshocks of quakes are regarded as the main contribution to significant disasters. To model the spatiotemporal patterns of earthquake shocks, we firstly consider the data attributes of specialty, which are discussed in Subsection~\ref{sec:data-cond}, then we pre-process the data in Subsection~\ref{sec:weaving}, and then the spatial and temporal dynamics are modeled in Subsection~\ref{sec:spatial},~\ref{sec:temporal} respectively.

The proposed prediction model consists of two major components, an autoencoder which learns the latent space distribution from the image-like view of the earthquakes, and a prediction network that learns to predict the likelihood of the next main shock happening within the same area.

\begin{figure*}[hpbt]
  \centering
  \centerline{\includegraphics[width=.88\linewidth]{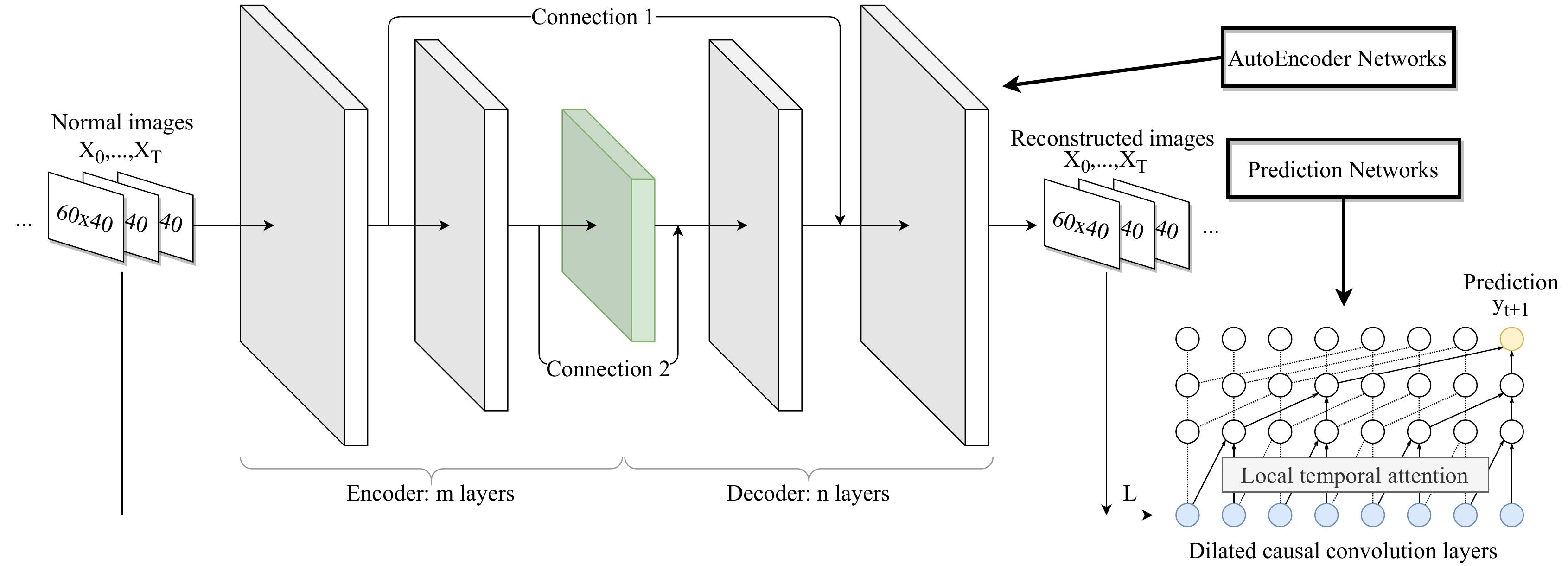}}
  \caption{\tseqpre: overview of earthquake prediction networks.}
  \label{fig:tseqpredictor}
\end{figure*}

\subsection{Energy-based Data Models}
\label{sec:data-cond}

Geographical data are coordinates related. Intuitively, those shocks that happened in different Geo terrain may take effect to future shocks unevenly.
The earthquake dataset used in this project is a tablet formatted catalog containing information on shocks in terms of geo-coordinates and magnitudes. In this project, we focus on the time and geolocation of shocks, other attributes like types of quakes are not included. This dataset covers all earthquake events in Southern California from the year 1990 to 2019. The full dataset is used in all the following experiments. Figure~\ref{fig:dataset} shows all events plotted in 2D maps, in which hot spots are areas where earthquakes frequently happened or big earthquakes happened in history. Figure~\ref{fig:mag-list} shows scatter plots of events with magnitudes equal and greater than 0, 2.5, 3.5, 4.5, respectively, where extreme cases are order-of-magnitude large quakes pre-defined by a picked threshold in this project according to the domain knowledge.

Seismometers record seismic events by calibrating the vibrations of waves. Magnitude in the dataset represents measured amplitude as a measured seismogram. While they are discrete data points, accumulating magnitudes by summing them up by averaging makes the temporal information loss and deemphasizes large earthquakes. In contrast to magnitude, earthquakes release energy can help mitigate this issue by two folds: 1) accumulated energy value in a region can represent the energy released by the stress of Earth over time; 2) energy data model naturally highlights large events since the energy of large events can be an order of magnitude higher than that of small events. The formula of converting earthquake magnitude to energy defines as:
\begin{equation}
  \mathbf{E} = (10^{\mathbf{Mag}})^{3/2}
\end{equation}
in which the magnitude $0 \leq \mathbf{Mag} \in \mathbb{R} \leq 10$.
Earthquake magnitude value can be even negative for tiny events that are negligible. This scale is also open-ended, but events with magnitude values greater than 10 are clipped to 10.

\subsection{Location-aware Data Weaving}\label{sec:weaving}
As a time-series prediction task, the earthquake catalog contains locations and magnitudes, which could be used as target properties. However, it could be more natural to reorganize the vector-valued 1D time-series dataset into a 2D sequence dataset by dividing a map region into small boxes according to longitudes and latitudes and aggregating the released energy within a small box per specific time-frequency.

Long short-term memory (LSTM)~\cite{hochreiter_long_1997} is an advanced model of recurrent neural networks suitable for modeling series-like vector-valued data.
Compared with the exiting LSTM approach such as~\cite{wang_earthquake_2020}, location-aware weaving gives finer-grained geolocations. Besides, 2D convolutional operations can put a strong prior on locations than the recurrent matrix multiplication for vector-valued observations in LSTM cells. Furthermore, for earthquakes, the underlying intuition is that 2D convolution may capture location-based plate movements, which is considered as the direct cause of earthquakes.

We denote $\mathbf{Mag^t_k}$ the value at location $k$ and time $t$ of a spatially and temporally continuous phenomenon of interest.
So each element of the sequence becomes
a summation of all energy released at the grid ($i,j$), given $i \in [0, M) $ and $j \in [0, N)$. Then the total energy for each grid element is defined in Eq~\ref{eq:summation}
, which means $X^t$ has a shape $M \times N$ for $M$ boxes along the latitude and $N$ boxes along the longitude.
\begin{equation}\label{eq:summation}
  X^t_{i,j} = \sum_{k \in grid[i][j]} ((10^{\mathbf{Mag^t_k}})^{3/2})
\end{equation}
Considering this area as a 2D mesh grid, this equation sums up all energy erupted in every grid for each time interval.
For example, we can sum up how much energy released within a box region of Longitude from -120 to -119 and Latitude from 32 to 33 everyday.

\subsection{Convolutional AutoEncoder for Effective Spatial Modeling}\label{sec:spatial}
Main shocks with large magnitudes are rare in terms of statistics and nature physics. In addition, earthquakes are full of stochastic processing, resulting in seismic signals are very noisy. To predict the future mainshocks, we first model the spatial patterns within the southern California area.

We use an autoencoder to mine the spatial pattern changes under normal circumstances and abnormal circumstances. As shown in Figure~\ref{fig:tseqpredictor}, the autoencoder consists of three major components: 1) a bunch of convolutional layers encodes the input to 2) the bottleneck layer in green color, and 3) the layers in decoder up-sample the latent variables from the bottleneck layer to the output. Compared to the variational autoencoder (VAE), we do not assume Gaussian distribution or any other distributions for the latent space. In addition, the reconstructed results from VAE are tended to be noisier. We also make some experiments for complete comparison in Section~\ref{sec:exp}.
Spatial modeling is a semi-supervised process to train a model that learns the representation of earthquake images. We train this model by minimizing the following equation.
\begin{equation}
  L(\mathbf{X_{normal}},g(f(\mathbf{X_{normal}})) + \Omega(h, \mathbf{X_{normal}})
\end{equation}
, where $ \mathbf{X_{normal}} $ are images of earthquakes with magnitudes less than a threshold, $f$ is an encoder function, $g$ is an decoder function, and $\Omega$ is a function that regularizes or penalizes the cost. This setup enforces the same input and output so that the bottleneck layer can obtain the most critical latent variables from the dataset. Detailed modeling methods used in AE are covered in the following sub-sections.

\subsubsection{Spatial modeling}
The encoder networks are comprised of convolutional layers followed by Batch Normalization with the ReLU activation function. The decoder networks are comprised of de-convolutional layers with the ReLU activation function.
After the seismic events are parsed and transformed to 2D image-like sequences in Section~\ref{sec:data-cond}, we can utilize the spatial dependencies between pixels. Convolutional operations are common image feature extraction means. Pixel relationships can be easily mapped to geology locations of events. The encoder component of this model is used to extract the spatial features from images which contribute to convolution layers in a downsampling manner. The downsampled feature maps enable the network to collect contextual information, which could be surface terrains on
Earth.  In convolution-base layers with a kernel $K$, the process takes the input $X$ with the following form:
\begin{equation}\label{eq:conv}
  S_{i,j} = \sum_m \sum_n X(i+m,j+n) K(m,n)
\end{equation}
The decoder component of this model is used for upsampling variables from the bottleneck layer by inverse 2D convolutional operations.
The final reconstructed image denoted as $\hat{X}$ is produced by the decoder.

\subsubsection{Skip connections}
We incorporate skip connections in the AutoEncoder architecture. Skip connections are forward shortcuts between layers in networks.
Skip connections can help recover the full spatial resolution at the network output to avoid the gradient vanishing, making fully convolutional methods suitable for modeling segments on maps.
They symmetrically connect layers from the encoder and decoder, as shown in Figure~\ref{fig:tseqpredictor}. This strategy allows long skip connections to pass features from the encoder path to the decoder path directly, which can recover spatial information lost due to downsampling, according to~\cite{he_identity_2016}. The combination of low-level features and high-level features improves the training performance due to large gradients and improves accuracy due to complementary information summarized from different levels.
Both long and short connections are used in the model as shown ``Connection 1'' and ``Connection 2'' in Figure~\ref{fig:tseqpredictor}.

The AE includes the bottleneck layer and other basic blocks. The design benefits of those are introduced in~\cite{he_deep_2016,he_identity_2016}. For instance~\cite{hou_deeply_2017} includes short skip connections for detecting salient objects.  Skipping some blocks with minimal modification encourages the information to pass through the non-linear functions to learn a residual representation from the direct input information.
Similar to the short skip connections in Residual Networks~\cite{long_fully_2015}, we sum the features from the encoder layers on the expanding path of decoder layers with long skip connections symmetrically.

\subsubsection{Bottleneck layer}
Hidden latent variables captured in the bottleneck layer have shown effectiveness in many applications, e.g.~\cite{li_diffusion_2017,dong_learning_2018}.
The bottleneck layer in the autoencoder is deliberately set to a small vector of a size $k$ feature map. This layer creates restrictions in the network to enforce the information pertaining to low dimensional space. Firstly, it regularizes the model from overfitting all samples. Secondly, a small feature map can better differentiate abnormal cases from normal cases. We set this $k$ as a hyperparameter in our model.

\subsection{Temporal Convolutional Model for Effective Temporal Modeling}\label{sec:temporal}

In this work, the goal of nowcasting earthquakes is to predict the future probability of the next major shock in Southern California. This can be done in a prediction network, which is fed in the information gained from the AutoEncoder. Typically a long short-term memory (LSTM) model can predict well on this task. However, in \tseqpre we incorporate an enhanced TCN (Figure~\ref{fig:tseqpredictor}), which can outperform LSTM. This situation is similar in predicting other physics-related fields of study. For example, TCN is used to predict climate changes~\cite{yan_temporal_2020}. We further analyze these features in the following sub-sections.

\subsubsection{Conditional Temporal Convolution}
Temporal convolutional neural networks are used to improve the temporal locality prediction over time. Temporal convolutional layers are layers containing causal convolution with varied dilation rate in 1D convolutional layers~\cite{oord_wavenet_2016,borovykh_conditional_2018}. A typical configuration of temporal convolutional layers is set the dilation rate corresponding to the i-th of layers, for example, $2^{i}$. Given a model with parameters $\theta$, the prediction is conditional on history events. This task can be expressed in the function:
\begin{equation}\label{eq:tcn}
  p(y|\theta)=\prod_{t=1}^{T}p(y_{t+1}|y_{1},\dots,y_{t},\theta)
\end{equation}

\subsubsection{Local Temporal Attention}
A localized attention process can enhance temporal information passing, which is inspired by self-attention structure from Transformer~\cite{vaswani_attention_2017}, and Hao \etal work for sequence modeling~\cite{hao_temporal_2020}. The process incorporates functions $f$, $g$, and $h$ to calculate $d$ dimensional vector of keys $\mathcal{K}$, queries $\mathcal{Q}$, and values $\mathcal{V}$ respectively. Then, we calculate the weight matrix by $W=\frac{\mathcal{K}\cdot\mathcal{Q}}{\sqrt{d}}$. Finally, we apply a softmax function to the lower triangle of $W$ to get a normalized attention weight $W_{attention}=softmax(W)$ and the final out of this layer can be calculated via this attention weighted summary: $\sum_{t=1}^{T}W_{attention}\cdot y_t$.

\subsubsection{Smooth Joint Nash–Sutcliffe Efficiency}
Nash–Sutcliffe model efficiency coefficient (NSE) is a commonly used metric to evaluate a predictive model. NSE is widely used to evaluate predictive skills in scientific studies, such as  hydrology~\cite{moriasi_model_2007}. The value range of NSE is $(-\infty, 1)$. NSE can become negative when the mean error in the predictive model is larger than one standard deviation of the variability. Its equation is defined as follows.
\begin{equation}
  NSE=1-\frac{\sum_{t=0}^{T}(\hat{y_t}-y_t)}{\sum_{t=0}^{T}(y_t-\Bar{y})}
\end{equation}
The goal of this metric is to force the predicted results to have a strong correlation between the distribution of predicted results and the distribution of expected results~\cite{mccuen_evaluation_2006}.

\subsection{Joint Probability Analysis}

Finally, we can combine the autoencoder and the temporal convolutional network together by connecting the reconstructed loss as shown in Figure~\ref{fig:tseqpredictor}.
After our model can learn the spatial dynamics via Eq~\ref{eq:conv} and temporal dynamics via Eq~\ref{eq:tcn}, the synthetic probability captured by the model can be defined as follows:
\begin{equation}
  \begin{split}
    Loss_{AE} &= MAE(X - \hat{X}) \\
    Loss_{TCN} &= NSE(y_t - \hat{y_t})
  \end{split}
\end{equation}
, which define the training targets.

In training predictive networks, we encode the output from AE and train in TCN. This process can be viewed as:
\begin{equation}
  \begin{split}
    y_t &= encode(X_t) \\
    \hat{y_t} &= TCN(y_i, y_{i+1}, \dots, y_{t-1})
  \end{split}
\end{equation}
, where $\hat{y_t}$ is directly related to the probability of the extreme events. This can be accessed by the defined threshold in AE.

\section{Experiments and Evaluation}\label{sec:exp}

The dataset is downloaded and parsed via the USGS website. It holds all earthquakes within the studied area in Southern California from 1990 to 2019. All the following experiments are conducted with the full dataset.

\subsection{Implementation Details and Experimental Setup}
Our \tseqpre model and other baseline models are implemented with TensorFlow 2~\cite{abadi_tensorflow_2016} in Python. We conduct experiments on a computer equipped with an Intel(R) Xeon(R) CPU E5-2670 v3 @ 2.30GHz, 128GB memory and 8 NVidia K80 GPUs. All models, including \tseqpre and baseline models, are trained using Adam or SGD optimizers with a fine-tuned learning rate and mean squared error as training loss. All model weights are check-pointed, and we select the best model weights for testing. Events with magnitudes $\geq 4.5$ are labeled as extreme major shocks.

The encoder consists of 2D convolutional layers by varying the filter size from 4, 16, 32, 64. And the layers in the decoder vary the filter size symmetrically.
The encoder and decoder are trained by minimizing the Mean Squared Error (MSE) loss between the input and its reconstruction.
We use the Adam optimizer with a learning rate of 0.001 by default. We set an early stop in the training process when the validation loss has stopped improving for 20 epochs and the best model is restored from checkpoints.
The training procedure iterates up to a maximum of 100 epochs. The batch size is set to 16, 64, and 128 respectively. In the temporal predictor, we test the time step size varying from 3 to 100 with appropriate filter sizes. The batch size is always set to one in order to make the model fully stateful.
Due to the stochastic nature of shocks, the output series from the autoencoder is denoised by the LOESS smoothing method~\cite{rojo_modeling_2017}.
We describe two groups of modeling approaches below: models in the first group have only one neural network, while the modeling from the second group uses a joint of two neural networks.
The modeling approaches compared in this project are listed as follows:
\begin{itemize}
  \item \modelnum{\bf MLP} refers to a neural network model with only fully connected dense layers. In TensorFlow, these layers are implemented with the ``keras.layers.Dense'' class. The number of layers in the MLP model is set as a hyperparameter. The number of neurons in the input and output layer are set according to the dataset. The number of neurons in the hidden layers varies from 32, 64, to 128. The following models labeled with `MLP' also apply this strategy.
  \item \modelnum{\bf LSTM} refers to a typical model with LSTM layers, including necessary input and output conversion. The recurrent unit in LSTM layers varies from 32, 64, to 128. These layers are implemented by the ``keras.layers.LSTM'' class.
  \item \modelnum{\bf Conv2D-FC} is a composite model in which 2D convolutional layers are followed by fully connected dense layers. Conv2D is brought by the ``keras.layers.Conv2D'' class in TensorFlow, while the filter size varies from 3 to 40 and kernel size varies from 7x7, 5x5, to 3x3.
  \item \modelnum{\bf Conv2D-LSTM} consists of one 2D convolutional layer to encode the input and one LSTM to predict the output.
  \item \modelnum{\bf ConvLSTM2D-FC} represents a model consisting of one ConvLSTM2D layer followed by a fully connected dense layer as the output. We use the TensorFlow class ``keras.layers.ConvLSTM2D'' as the implementation of the ConvLSTM2D layer, the algorithm of which was introduced from~\cite{shi_convolutional_2015}. The hyperparameters of ConvLSTM2D are similar to those used in Conv2D and LSTM. FC layer is set as above.
  \item \modelnum{\bf MLP+MLP} refers to joint training of two MLP models. The first MLP refereed as a dense AE consisting of 3 to 5 dense layers encoding and decoding the input. The second MLP represents a 3-layer fully connected neural network for prediction.
  \item \modelnum{\bf MLP+LSTM} refers to joint modeling with an MLP model and an LSTM model as mentioned above, where an MLP model acts as the AE and an LSTM model acts as the predictor.
  \item \modelnum{\bf MLP+Conv1D} is similar to the above case, except that one Conv1D-based model is used as the predictor, in which the filter size is set according to the task and the kernel size varies from 3 to 50 continuously.
  \item \modelnum{\bf Conv2D+MLP} combines a model with Conv2D layers as the AE and a simple MLP model as a predictor. Hyperparameters used in the two models are set similarly to those mentioned above.
  \item \modelnum{\bf Conv2D+LSTM} uses a Conv2D-based model and a LSTM-based model defined as above. Hyperparameters used in the experiments are also similar.
  \item \modelnum{\bf Conv2D+Conv1D} combines a model with Conv2D layers as the AE and a Conv1D-based model as the predictor. Hyperparameters used in the experiments are also similar.
\end{itemize}

We train the models \circled{1} to \circled{5} with the Adam optimizer and the NSE loss. For the above methods \circled{6} to \circled{11} and our \tseqpre, the autoencoder is trained with the Adam optimizer and an MSE loss. The prediction network is trained with the Adam optimizer and an NSE loss.
While many hyperparameters are required in all these experiments, we tend to choose the best set of parameters for each set of models and prevent them from overfitting.

\subsection{Dataset Preprocessing and Augmentation}
The earthquake dataset is a tablet formatted dataset in which each record is an earthquake epicenter with a timestamp, a GEO location, a magnitude, and depth. We pre-process the events from the catalog by accumulating the events and parsed as spatial grids according to the analysis in Section~\ref{sec:data-cond}. We implement the data feeding pipeline with the TensorFlow data streaming APIs. The prepared dataset is loaded from disks with multiple workers and then fed in the training model from CPU to GPU batch by batch so that the pipeline can boost the I/O parallelism and increase the CPU and GPU efficiency.

\begin{figure}
    \begin{minipage}[b]{0.48\linewidth}
        \centering
        \includegraphics[width=\linewidth]{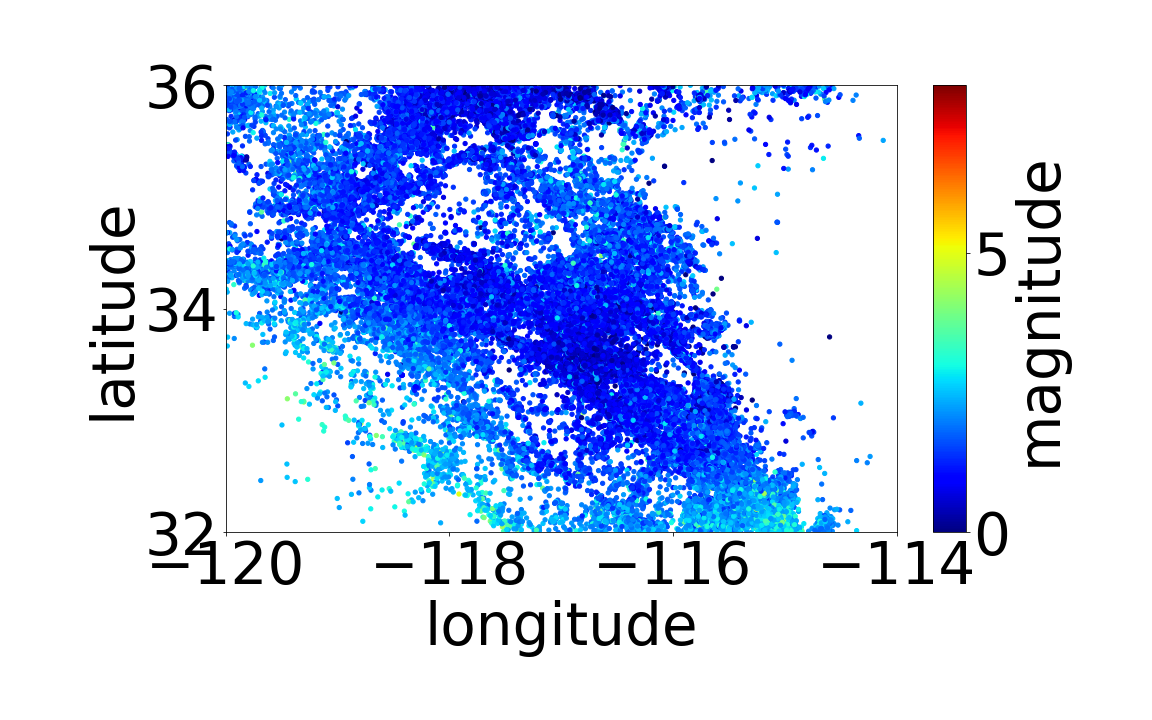}
        \centerline{(a) Mag $>=0$}
    \end{minipage}
    \hfill
    \begin{minipage}[b]{0.48\linewidth}
        \centering
        \includegraphics[width=\linewidth]{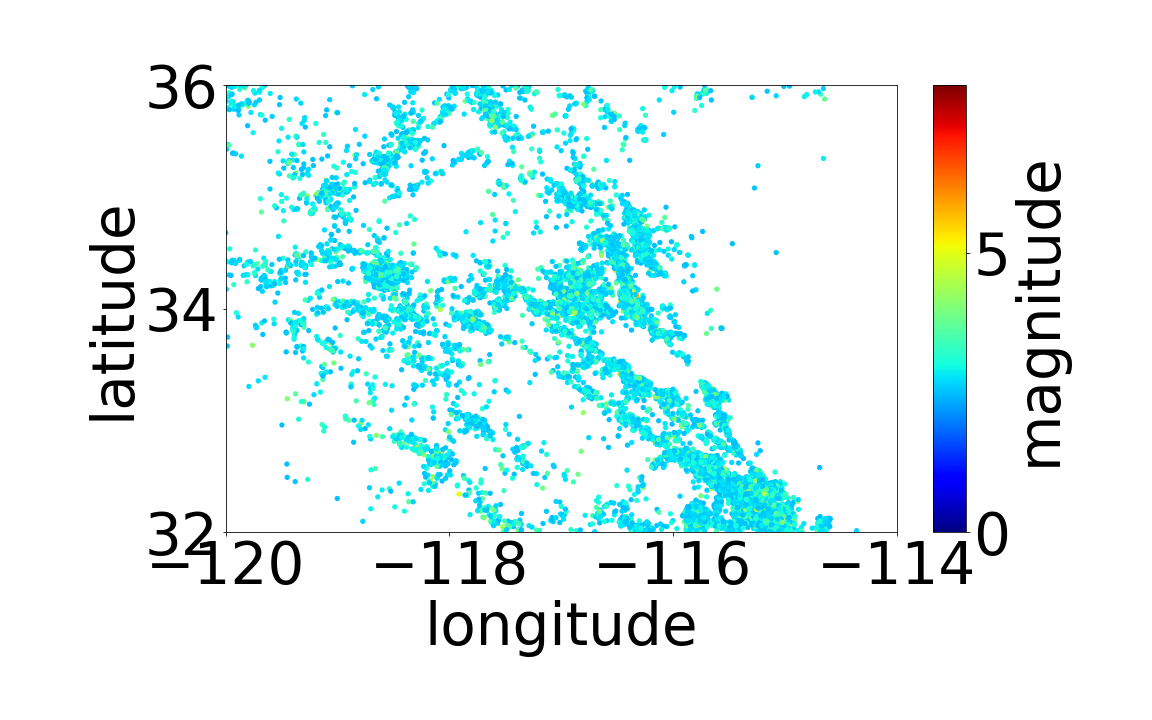}
        \centerline{(b) Mag $>=2.5$}
    \end{minipage}
    \begin{minipage}[b]{0.48\linewidth}
        \centering
        \includegraphics[width=\linewidth]{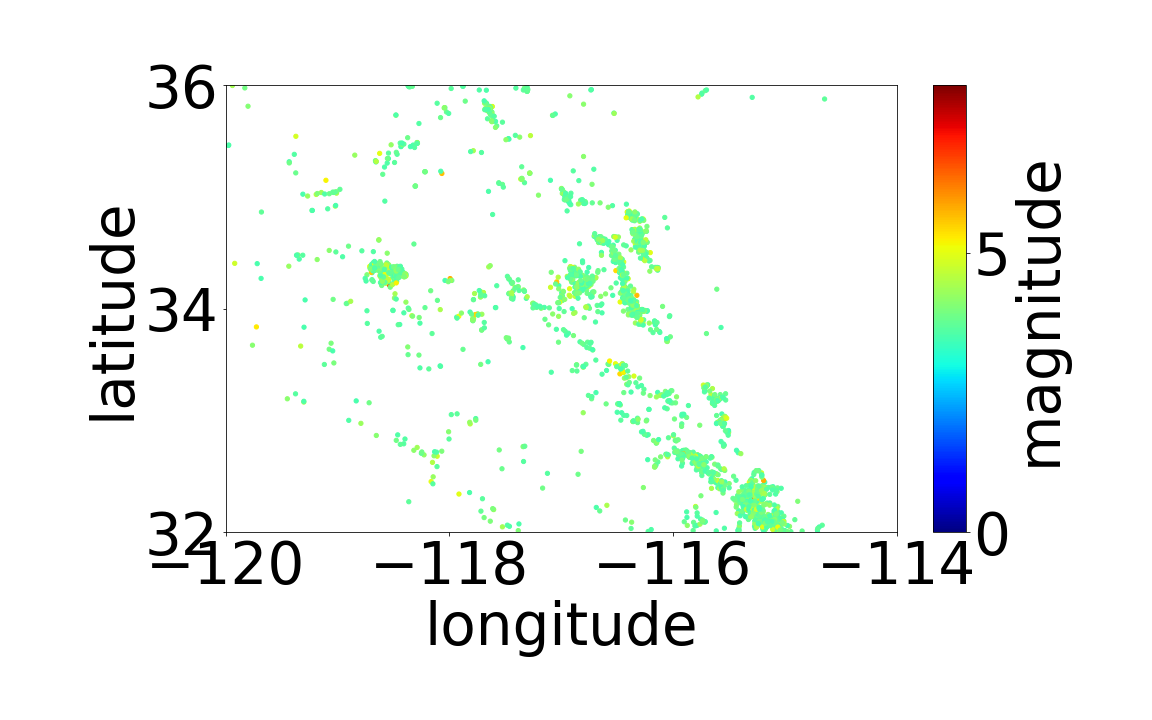}
        \centerline{(c) Mag $>=3.5$}
    \end{minipage}
    \hfill
    \begin{minipage}[b]{0.48\linewidth}
        \centering
        \includegraphics[width=\linewidth]{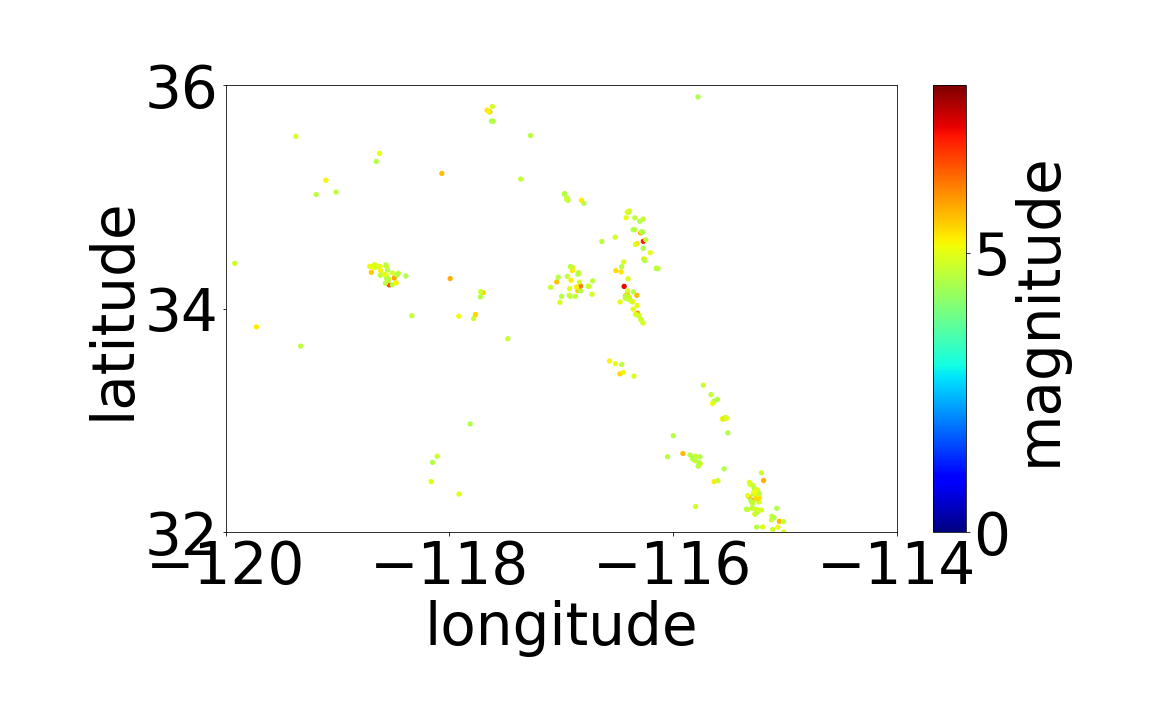}
        \centerline{(d) Mag $>=4.5$}
    \end{minipage}
    \caption{Dataset overview: (a) $444,589$ events with magnitude $\geq 0.0$, (b) $24,822$ events with magnitude $\geq 2.5$, (c) $2,489$ events with magnitude $\geq 3.5$, (d) $237$ events with magnitude $\geq 4.5$. We can observe that the number of larger earthquakes is an order of magnitude less than smaller ones.}
    \label{fig:mag-list}
\end{figure}

Figure~\ref{fig:mag-list} shows four scatter plots by filtering events with magnitudes greater than or equal to 0, 2.5, 3.5, 4.5 in (a), (b), (c), and (d) respective sub-figures.
We divide the Southern California (Longitude: -120\textdegree{}\textasciitilde-140\textdegree{}, Latitude: +32\textdegree{}\textasciitilde+36\textdegree{}) into a grid with $60\times 40$ cells, each of which has 0.1 degree of longitude and latitude for about 11.1km (1 degree in kilometers is about 111km). Firstly, it is easy to group events into daily intervals. Then, let $x$, $y$ denote the longitude and latitude location of an event. All events are accumulated in the corresponding cell where $x$, $y$ fall into. The value of each cell is the mean of magnitudes of all events within the cell. As a result, each day is represented by a 2D image-like $60\times 40$ matrix.


\subsection{Performance Analysis}\label{sec:perf-analysis}

In these sets of experiments, we aim to demonstrate the performance of \tseqpre compared to a series of baseline models. Firstly, we show the performance differences between autoencoder in \tseqpre and a VAE. Then, we compare the prediction network with a LSTM. Finally, we illustrate the comprehensive results from using \tseqpre comparing with a series of methods.

\begin{table*}
    \centering
    \caption{Results comparison between \tseqpre and baseline models. Some models adopt the same architecture of using an autoencoder and a prediction network. These models are named with a `$+$' sign. Please note some models do not have meaningful MAE values, which are omitted in the table. Please refer to Section~\ref{sec:perf-analysis} for more details.}
    \label{tbl:overall}
    \begin{tabular}{|l|r|r|r|r|r|r|}
        \hline
        \textbf{Models}   & \multicolumn{1}{l|}{\textbf{MAE}} & \multicolumn{1}{l|}{\textbf{Precision}} & \multicolumn{1}{l|}{\textbf{Recall}} & \multicolumn{1}{l|}{\textbf{F-1}} & \multicolumn{1}{l|}{\textbf{F-0.2}} & \multicolumn{1}{l|}{\textbf{NSE}} \\ \hline \hline
        \circled{1} MLP               & \multicolumn{1}{l|}{-}            & 0.2631                                  & 0.2845                               & 0.2096                            & 0.2494                              & -1.4739                           \\ \hline
        \circled{2} LSTM              & \multicolumn{1}{l|}{-}            & 0.4596                                  & 0.5186                               & 0.3801                            & 0.4058                              & -0.2059                           \\ \hline
        \circled{3} Conv2D-FC         & \multicolumn{1}{l|}{-}            & 0.4589                                  & 0.3963                               & 0.4340                            & 0.4394                              & -0.1867                           \\ \hline
        \circled{4} Conv2D-LSTM       & \multicolumn{1}{l|}{-}            & 0.4299                                  & 0.4069                               & 0.4217                            & 0.4243                              & -0.4022                           \\ \hline
        \circled{5} ConvLSTM2D-FC     & \multicolumn{1}{l|}{-}            & 0.4633                                  & 0.3289                               & 0.3763                            & 0.3801                              & -0.1714                           \\ \hline
        \circled{6} MLP+MLP           & 0.2570                            & 0.7525                                  & 0.6338                               & 0.6652                            & 0.7113                              & 0.6778                            \\ \hline
        \circled{7} MLP+LSTM          & 0.1637                            & 0.8420                                  & 0.7085                               & 0.7599                            & 0.8021                              & 0.7890                            \\ \hline
        \circled{8} MLP+Conv1D        & 0.1484                            & 0.8571                                  & 0.9351                               & 0.8029                            & 0.8342                              & 0.8133                            \\ \hline
        \circled{9} Conv2D+MLP        & 0.1484                            & 0.8577                                  & 0.7944                               & 0.7887                            & 0.8098                              & 0.8108                            \\ \hline
        \circled{10} Conv2D+LSTM       & 0.1410                            & 0.8640                                  & 0.8776                               & 0.8609                            & 0.8683                              & 0.8222                            \\ \hline
        \circled{11} Conv2D+Conv1D     & 0.0588                            & 0.9420                                  & 0.9115                               & 0.8998                            & 0.8688                              & 0.9293                            \\ \hline
        \textbf{\tseqpre} & \textbf{0.0483}                   & \textbf{0.9563}                         & 0.9016                               & \textbf{0.9251}                   & \textbf{0.9341}                     & \textbf{0.9323}                   \\ \hline
    \end{tabular}
\end{table*}

\subsubsection{AutoEncoder}

We use Mean Absolute Error (MAE) as a metric to test the AE performance, the results of which are listed in the MAE column from Table~\ref{tbl:overall}. Models from \circled{1} to \circled{6} do not contain AE, so the results of them are omitted in the table. From this table, the MAE score of our \tseqpre is 0.0483 and the lowest score for other models is 0.0588. \tseqpre outperforms all other modeling approaches from \circled{6} to \circled{11}, whereases Conv2D-based autoencoders can generate competitive performance. From another aspect, this means the convolutional layers actually work for mining spatial patterns.

\begin{table}
    \centering
    \caption{\tseqpre AutoEncoder vs. VAE.}
    \label{tbl:auto-vs-vae}
    \begin{tabular}{|l|c|c|c|}
        \hline
        \textbf{Model}                       & \textbf{MSE} & \textbf{Accuracy} & \textbf{Variance} \\ \hline \hline
        \tseqpre                             & 0.148        & 0.968             & 1.432             \\ \hline
        VAE~\cite{kingma_auto-encoding_2013} & 0.157        & 0.971             & 1.986             \\ \hline
    \end{tabular}
\end{table}

We also compare the autoencoder used in \tseqpre in Section~\ref{sec:tseqpredictor} as opposed to a variational autoencoder (VAE) separately. We test some metrics of using \tseqpre autoencoder and a common VAE. The performance results are summarized in Table~\ref{tbl:auto-vs-vae}, from which our \tseqpre outperforms a single VAE. Even though VAE can achieve almost the same performance in terms of accuracy, it has a higher mean squared error loss and variance for the final output. Higher MAE loss and variance affect the performance of the prediction network. Besides the benefits from applying skip connections may outweigh other aspects for this case.

\begin{table}
    \centering
    \caption{Varying the latent space dimension.}
    \label{tbl:latent-dim}
    \begin{tabular}{|l|l|l|}
        \hline
        \textbf{Latent space dimension} & \textbf{MSE} & \textbf{Accuracy} \\ \hline \hline
        16                              & 0.148        & 0.968             \\ \hline
        64                              & 0.140        & 0.968             \\ \hline
        128                             & 0.138        & 0.972             \\ \hline
        1024                            & 0.137        & 0.984             \\ \hline
    \end{tabular}
\end{table}

To show the critical hyperparameter that affects the overall performance, we list the results varying the latent space dimension from 16 to 1024, as shown in Table~\ref{tbl:latent-dim}. Larger latent space tends to have better fitting to the dataset, however this value should be regularized as small as possible.

\subsubsection{Prediction}

We use four metrics to compare all temporal predictors from \circled{1} to \circled{11} with \tseqpre: Precision, Recall, F-1, F-0.2, and NSE, where precision and recall are used to evaluate the capability of predicting positive events and F-scores give overview of accuracy.
Even though basic models from \circled{1} to \circled{5} can predict some events. Their NSE scores are very small, which means those models are not reliable. Modeling \circled{6} to \circled{11} can have competitive scores. \circled{11} can have a better Recall value than \tseqpre.
In contrast to modeling approaches from \circled{1} to \circled{11}, the prediction network in \tseqpre can outperform all these approaches overall. We summarize the experimental results in Table~\ref{tbl:overall}.

\subsubsection{Comprehensive Analysis}

In this set of experiments, we list several commonly used models for predicting the future main shocks. The results are summarized in Table~\ref{tbl:overall}. In this table, MLP represents a three-layer of fully connected neural networks. LSTM represents a two-layer of stateful LSTM neural networks. Conv2D, Conv1D represent a neural network consisting of one 2D convolutional and one 1D convolutional layer, respectively. From this table, we illustrate \tseqpre can outperform a single model significantly and other combination of models for this task.
Please note the classes are not balanced in this case, so that F-1 and F-0.2 scores may be higher than expected but we keep them as a reference here.





\subsection{Ablation Studies}

To verify the effectiveness of skip connections and local temporal attention applied in \tseqpre, we test the models and compare the performance without these techniques.
In the following two sets of experiments, we demonstrate the two primary techniques that can improve the autoencoder and the prediction network: skip connections and local temporal attention. In the first set, we remove the skip connections in the autoencoder and keep the remaining parts the same. In the second set, we remove the local temporal attention in the prediction network and use the same autoencoder as the \tseqpre. Table~\ref{tbl:ablation} shows the results of these two sets of experiments.

\begin{table}
    \centering
    \caption{Ablation study by removing core components in \tseqpre.}
    \label{tbl:ablation}
    \begin{tabular}{|l|l|l|}
        \hline
        \textbf{Models}              & \textbf{F-1}    & \textbf{NSE}    \\ \hline \hline
        W/O skip connections         & 0.9001          & 0.9233          \\ \hline
        W/O local temporal attention & 0.9247          & 0.9289          \\ \hline
        \tseqpre (with both)         & \textbf{0.9251} & \textbf{0.9323} \\ \hline
    \end{tabular}
\end{table}

\subsection{Discussion and Empirical Study}

\begin{figure}
  \centering
  \includegraphics[width=.78\linewidth]{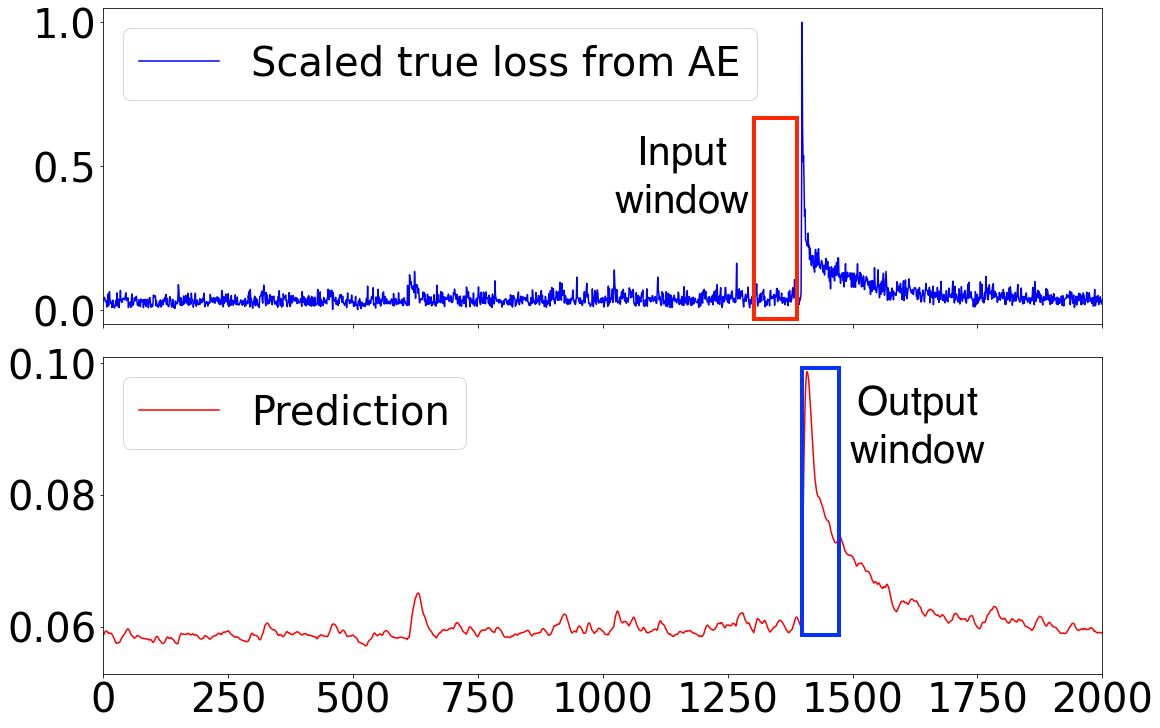}
  \caption{\tseqpre prediction.}
  \label{fig:prediction}
\end{figure}

We build joint models as shown in Figure~\ref{fig:tseqpredictor}, in which the autoencoder can learn the spatial pattern and the predictor can forecast future events.
Figure~\ref{fig:prediction} shows a prediction example. Given an input sequence window, the predictor can output a future sequence window, from which a major shock can be detected.
There are two aspects in the consideration of this model: 1) During the training period, a sequence of $T$ 2D matrices collected from time $t_1, t_2, \ldots$ to $t_T$ are the input: $X_{t_1}, X_{t_2},\ldots,X_{t_T}$, and the output is another sequence:$y_{t_2}, y_{t_3},\ldots, y_{t_{T+1}}$. In this way, the $y_{t_{T+1}}$ is the predicted result. This means that the model can be trained on rolling basis as the data stream in.
2) In Southern California, the model can be trained and predict a novelty score representing the probability of the next major shock. For example, if the input is $X_t$ at $t$ time, the output from the model is $X_{t+1}$ at $t+1$ time. The predicted probability of this area can be told from $y_{t+1}$.

This modeling approach and experimental results are still empirical.
Throughout all experiments, according to Figure~\ref{fig:mag-list}, we test pre-defined thresholds that filter extremely large quakes of magnitude at 4, 4.2, and 4.5. In all cases, this modeling approach can generalize to the same effective results. However, we have not expanded the dataset to cover a larger area in Southern California or a different area on the earth.



\section{Conclusions and Future Work}
\label{sec:con}

In this project, we propose \tseqpre, a joint modeling approach that mines the spatial and temporal dynamics from the dataset and predict extreme event by using learned latent variables. We dissect the problem settings for forecasting earthquakes, discuss how we model spatial temporal forecasting problems using deep neural networks.
In contrast to 11 different approaches in the experiments, we demonstrate the effectiveness of \tseqpre to predict extreme cases in Southern California. According the metrics from our experiments, we show some promising when proper thresholds are chosen to filter out noisy.
Even though we have study a few modeling approach and find our the most effective one, the domain knowledge is still required from Geoscience experts. In future, we plan to verify this approach in wider areas and we also consider other physics quantities like seismicity, electric field, magnetic field, deformation which are highly possible correlated to earthquake events.

\subsubsection*{Code and data availability}
The earthquake raw event dataset used in the paper is available to download from the USGS website at \url{https://www.usgs.gov/}. Model codes and parsed datasets used in the paper will be published at \url{https://github.com/DSC-SPIDAL/TS-earthquake}.

\section*{Acknowledgement}
This work is partially supported by the National Science Foundation (NSF) through awards CIF21 DIBBS 1443054, SciDatBench 2038007, CINES 1835598, and Global Pervasive Computational Epidemiology 1918626. We are grateful to Cisco University Research Program Fund grant 2020-220491 for supporting this research. We thank the Futuresystems team for their support.

\bibliographystyle{IEEEtran}
\bibliography{zotero}

\end{document}